\documentclass[%
 reprint,
nofootinbib,
 amsmath,amssymb,
 aps,
floatfix,
]{revtex4-2}

\usepackage{graphicx}
\usepackage{dcolumn}
\usepackage{appendix}
\usepackage{longtable}
\usepackage{color}
\DeclareSymbolFontAlphabet{\amsmathbb}{AMSb}\usepackage{amsthm}
\DeclareMathAlphabet{\mathpzc}{OT1}{pzc}{m}{it}
\usepackage{bm}
\usepackage{hyperref}

\begin{document}


\title{Universality of minimal length}
\author{Ahmed Farag Ali $^\nabla$$^{\triangle}$}
\email[email: ]{afaragali@ucmerced.edu; ahmed.ali@fsc.bu.edu.eg}
\author{Ibrahim Elmashad $^{\triangle}$}
\email[email: ]{ibrahim.elmashad@fsc.bu.edu.eg}
\author{Jonas Mureika$^{\square, \bigcirc }$}
\email[email: ]{jmureika@lmu.edu}
\affiliation{ $^{\triangle}$ Dept. of Physics, Benha University, Benha 13518, Egypt}
\affiliation{{$^\nabla$ Dept. of Physics, University of California Merced, 5200 North Lake Road, Merced,
CA 95344, United States}}
\affiliation{ $^{\square}$ Department of Physics, Loyola Marymount University, 1 LMU Drive, Los Angeles, CA, USA 90045
}
\affiliation{ $^{\bigcirc}$ Kavli Institute for Theoretical Physics, University of California Santa Barbara, Santa Barbara, CA}

\begin{abstract}
We present an argument reinterpreting the generalized uncertainty principle (GUP) and its associated minimal length as an effective variation of Planck constant ($\hbar$), complementing Dirac's large number hypothesis of varying $G$. We argue that the charge radii ({\it i.e.} the minimal length of a scattering process) of hadrons/nuclei along with their corresponding masses support an existence of an effective variation of $\hbar$. This  suggests a universality of a minimal length in measurement of scattering process. Varying $\hbar$ and  $G$ explains the necessity of Von Neumann entropy correction in Bekenstein-Hawking entropy-area law. Lastly, we suggest that the effective value of $\hbar$ derived from various elements may be related to the epoch of their creation via nucleosynthesis.\\
\begin{center}
{\bf``Change is the only constant'' Heraclitus}
\end{center}
\end{abstract}

\maketitle


\maketitle

\noindent {\bf{\Large{V}}}arious approaches to quantum gravity such as string theory, loop quantum gravity  and quantum geometry suggest an existence of a minimal measurable length. These features are collectively known as the generalized uncertainty principle (GUP)\cite {Amati:1988tn,Garay:1994en,Scardigli:1999jh,Brau:1999uv,Konishi:1989wk,Kempf:1994su,Maggiore:1993rv,Capozziello:1999wx,Ali:2009zq,Das:2010zf,Isi:2013cxa,Zhu:2008cg,Mignemi:2009ji,Bishop:2018mgy,Mureika:2018gxl,Knipfer:2019pgi,Pedram:2011gw,Shababi:2017zrt,Fadel:2021hnx,Lambiase:2022xde}. One of its renowned forms is given by:

\begin{equation}
\label{gupquadratic}
\Delta x\Delta p \geq \frac{\hbar}{2}(1+\beta\Delta p^2),
\end{equation}
where $\beta=\beta_0 \ell_p^2/\hbar^2$, $\beta_0$ is a dimensionless constant, and $\ell_p=1.6162\times 10^{-35}\text{m}$ is the Planck length. There have been several phenomenological and experimental studies on the generalized uncertainty principle in low and high energy systems such as atomic systems \cite{Das:2008kaa}, quantum optical systems \cite{Pikovski:2011zk}, gravitational bar detectors \cite{Marin:2013pga}, gravitational decoherence \cite{Petruzziello:2020wkd}, composite particles \cite{Kumar:2019bnd}, astrophysical systems \cite{Moradpour:2019wpj,Vagnozzi:2022moj}, inflationary models of the universe \cite{Easther:2001fi},  condensed matter systems \cite{Iorio:2017vtw}, baryon asymmetry \cite{Das:2021nbq} and macroscopic harmonic oscillators \cite{Bawaj:2014cda}. This array of results suggest considerably different bounds on the GUP parameter, and thus the corresponding minimal length \cite{Ali:2011fa,Pikovski:2011zk}. Several forms of GUP have been proposed \cite{Amati:1988tn,Garay:1994en,Scardigli:1999jh,Brau:1999uv,Konishi:1989wk,Kempf:1994su,Maggiore:1993rv,Capozziello:1999wx,Ali:2009zq,Das:2010zf,Isi:2013cxa,Zhu:2008cg,Mignemi:2009ji,Bishop:2018mgy,Mureika:2018gxl,Knipfer:2019pgi,Pedram:2011gw,Shababi:2017zrt,Fadel:2021hnx}. Collectively, these differing GUP models can be written in the following form$:$

\begin{equation}
    \Delta x~ \Delta p ~\geq ~\frac{\hbar}{2}~ f(p,x), \label{GUP}
\end{equation}
where $f(p,x)$ represents a generic modification of the uncertainty principle that depends on momentum in some models \cite{Amati:1988tn,Garay:1994en,Scardigli:1999jh,Brau:1999uv,Konishi:1989wk,Bishop:2018mgy,Capozziello:1999wx,Ali:2009zq,Das:2010zf}, and on position in others \cite{Zhu:2008cg,Mignemi:2009ji,Mureika:2018gxl}. A review of the GUP and its phenomenological and experimental implications can be found in \cite{Hossenfelder:2012jw}. An original perspective can be gained from Equation~(\ref{GUP}), however, by redefining $\hbar f(p,x)$ as an {\it effective Planck constant} $\hbar^{\prime}$, and thus the GUP becomes:
\begin{equation}
    \Delta x~ \Delta p ~\geq ~\frac{\hbar^{\prime}}{2}, \label{varyh}
\end{equation}
Note $\hbar^{\prime}$ still implies the existence of a minimal length. The quadratic form of GUP \cite {Amati:1988tn,Garay:1994en,Scardigli:1999jh,Brau:1999uv,Konishi:1989wk,Kempf:1994su,Maggiore:1993rv,Capozziello:1999wx} was conceptualized as an effective variation of Planck constant in \cite{Chang:2001bm,Chang:2011jj} that lead to an invariant phase space under time evolution in the context of  Liouville theorem. In this letter, we are not considering which model or functional form of $\hbar^{\prime}$ is more suitable for measurements. We are instead interested in a deeper question:{\it What is the reason for believing that $\hbar$ varies?}. To answer this question, let us start with the definition of $\hbar$ in terms of the fundamental Planck units, {\it i.e.}

\begin{equation}
    \ell_{P} ~ M_{P}~ c= \hbar,  \label{planck}
\end{equation}
where $c$ is the speed of light, and $\ell_{P}$ and $M_{P}$ are the Planck length and Planck mass, respectively. The latter two quantities represent a {\it minimal length}
and  {\it maximal mass} imposed by Nature.
These unique scales do not exist mathematically in relativistic quantum theories. This problem has been resolved in doubly special relativity (DSR) that was proposed by Magueijo and Smolin \cite{Magueijo:2001cr}. DSR suggests the existence of an invariant length/energy scale in addition to the invariance of the speed of light. The corresponding uncertainty principle of doubly special relativity was introduced and investigated in \cite{Ali:2009zq,Cortes:2004qn} and it was found that it implies discreteness of space \cite{Ali:2009zq,Das:2010zf}, which can also be represented by the generic Equation (\ref{varyh}). This model of GUP is written as follows:
\begin{eqnarray}
\label{gup}
[x_i, p_j] = i \hbar\hspace{-0.5ex} \left[  \delta_{ij}\hspace{-0.5ex}
- \hspace{-0.5ex} \alpha\hspace{-0.5ex}  \left( p \delta_{ij} +
\frac{p_i p_j}{p} \right)
+ \alpha^2 \hspace{-0.5ex}
\left( p^2 \delta_{ij}  + 3 p_{i} p_{j} \right) \hspace{-0.5ex} \right],~~~
\end{eqnarray}
where $\alpha=\alpha_0 l_p/\hbar$, and $\alpha_0$ is a dimensionless constant.

\begin{figure}
    \centering
    \includegraphics [width= 9 cm]{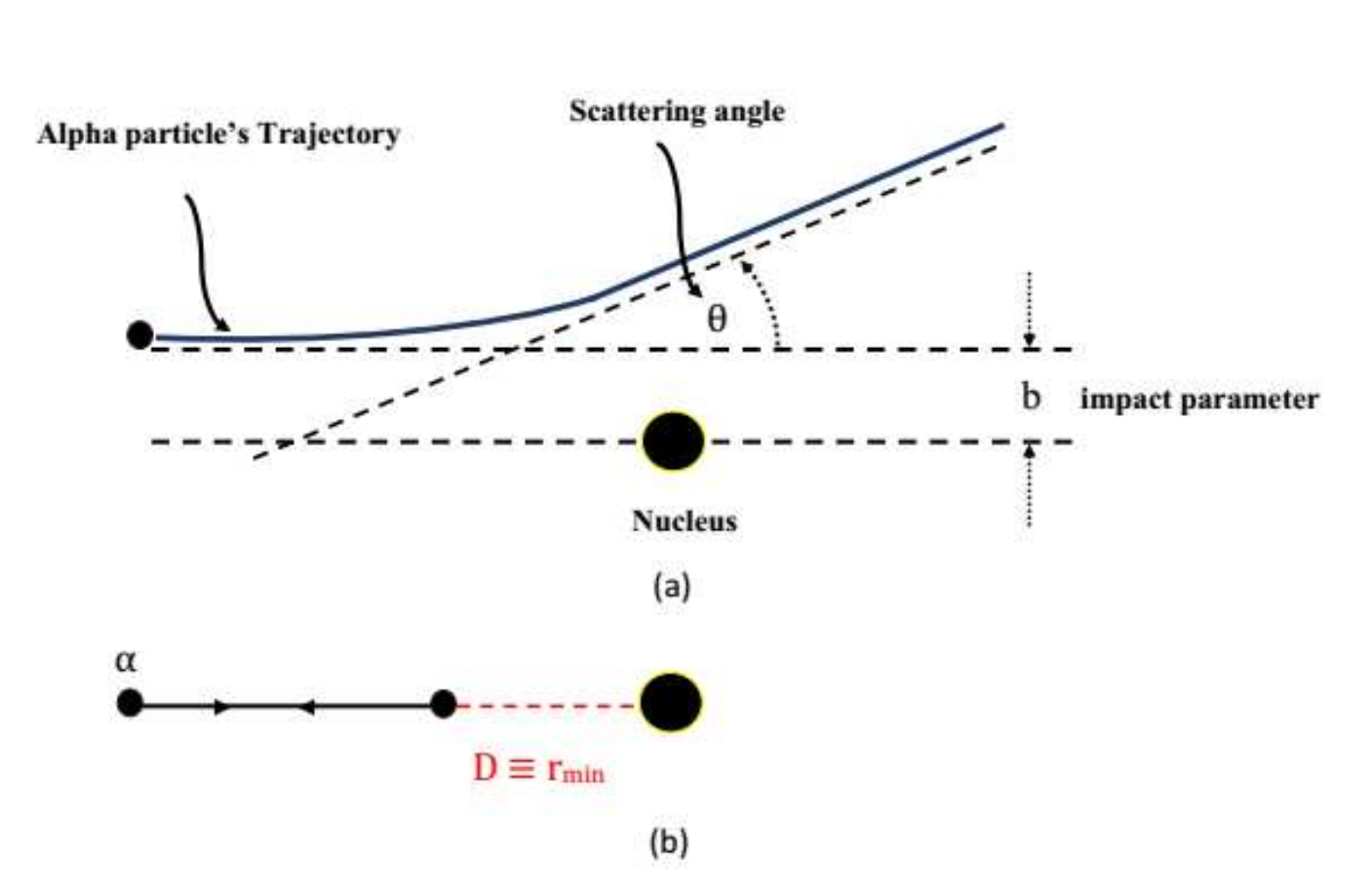}
    \caption{(a) Rutherford scattering of the $\alpha-$particle scattering on a very thin gold foil. (b) The distance of the closest approach $D$ at $\theta= \pi$.}
    \label{minimallength}
\end{figure}

We now seek to establish a version of Equation~(\ref{planck}), which defines a constant $\hbar$, that incorporates Equation~(\ref{varyh}) to define a varying $\hbar$. One can imagine the Planck length as the radius of a fundamental ``nucleus'' from which every particle is scattered. This interpretation is similar to Rutherford's historic discovery of the nucleus by analysing $\alpha-$particle scattering on a very thin gold foil \cite{rutherford1911lxxix} using the data of Geiger and Marsden (as shown in Figure~\ref{minimallength}(a)). The gold nucleus was treated as a point particle that is sufficiently massive relative to the mass of incident $\alpha-$particle, so that any nuclear recoil could be ignored. The key idea in this experiment was the existence of a { \it specific distance of closest approach}, $D$, at which the $\alpha-$particle is obligated to completely reverse its direction ({\it i.e.} the scattering angle $\theta$ equals $\pi$ as shown in Figure~\ref{minimallength}(b)). This specific distance can be obtained by equating the Coulomb energy with the initial kinetic energy of $\alpha-$ particle. In the context of quantum gravity, the analogue of this distance is a minimal length that may be similar to the Planck length $\ell_{P}$ but at different energy scale. The scattering relation is $:$
\begin{equation}\label{Ruth1}
\tan\left(\dfrac{\theta}{2}\right)= \dfrac{D}{2 b},
\end{equation}
where $b$ is the impact parameter.
In atomic, nuclear, and particle physics, the concept of charge radius measures the minimal length size of a composite hadron or nuclei. For example, a charge radius has been measured for the proton \cite{Pohl:2010zza,Sick:2003gm,Antognini:2013txn}, neutron \cite{Kopecky:1995zz,Foldy:1958zz}, pion \cite{Maris:1999bh}, deuteron \cite{Sick:1998cvq} and kaon\cite{KTeV:2005eic}. In that sense, we can set a generalization of Equation~(\ref{planck}) in terms of the charge radius and mass of the corresponding system. We take the same form of Equation~(\ref{planck}), replacing Planck length $\ell_P$ by the charge radius ($r$) \cite{Xiong:2019umf} and the Planck mass $M_{P}$ by the mass of the hadron/nuclei ($m$). This would take the following form$:$
\begin{eqnarray}
r~m~c = \hbar^{\prime},
\end{eqnarray}
where $m$ is the particle's mass and $\hbar^{\prime}$ is the effective Planck's constant. Here, $c$ is chosen to be constant to maintain consistency with the theory of relativity. For the proton, the charge radius is $r_p=0.831$~fm \cite{Xiong:2019umf} and the mass is $m_p= 1.672 \times 10^{-27}$ kg
 \cite{ParticleDataGroup:2020ssz}. Surprisingly, we obtain a value for $\hbar^\prime$ that is around four times the accepted value of Planck constant's, {\it i.e.}
\begin{eqnarray}
\text{Proton case:}~~~r_p~m_p~c= 3.95~\hbar,
\end{eqnarray}

Let us repeat the same computations for the pion $\pi$, which has charge radius $r_\pi= 0.657 \pm 0.003 $ fm \cite{Maris:1999bh,Ananthanarayan:2017efc} and mass $m_{\pi}= 2.488\times  10^{-28}$ kg \cite{ParticleDataGroup:2020ssz}:
\begin{eqnarray}
\text{Pion case:}~~~r_{\pi}~m_{\pi}~c=   0.449 ~\hbar,
\end{eqnarray}
Here, the result obtained is of the order of magnitude of Planck's constant. In the case of the Kaon, where $r_K=0.56 \pm 0.031$ fm \cite{Amendolia:1986ui} and $m_K=8.8\times10^{-28}$ kg \cite{ParticleDataGroup:2020ssz}:
\begin{eqnarray}
\text{Kaon case:}~~~~r_{k}~m_{k}~c=   1.458 ~\hbar,
\end{eqnarray}
Once again, the result we obtained is roughly equal to Planck's constant. Lastly for the Deuteron \cite{Sick:1998cvq} $r_D=2.130\pm 0.003$ fm, and  $m_D = 3.343 \times 10^{-27}$ \cite{Sick:1998cvq,ParticleDataGroup:2020ssz}, we find:
\begin{eqnarray}
\text{Deuteron case:}~~~~r_D~m_D~c=  20.22~\hbar,
\end{eqnarray}
which is larger than Planck's constant by a factor of 20.

We note that the effective Planck's constant $\hbar^\prime$ fluctuates around the true value of Planck's constant by few orders of magnitude, which may suggest that effective value of $\hbar$ is actually dependent on the given physical system. This is in fact consistent with the GUP Equation~(\ref{GUP}), and implies that GUP parameter bounds would be varying for different physical systems\cite{Ali:2011fa,Das:2008kaa}. We conclude that a particle's charge radius and the mass introduce a consistent picture between Equation~(\ref{varyh}) and Equation~(\ref{planck}).

 Furthermore, a recent result that may support this variation is the direct measurements of a varying fine structure constant \cite{Wilczynska:2020rxx} due to gravitational effects. It is well-known that the wide array of GUP models are motivated by different approaches to quantum gravity and each introduce correspondingly different corrections to quantum systems. It is therefore logical to suggest that $\hbar$ is varying due to the same quantum gravity effects. Moreover, a recent study sets bounds on varying $\hbar$ in terms of violation of local position invariance  \cite{PhysRevLett.108.110801}.

\begin{figure}[htp]
\centering
\includegraphics[scale=0.4]{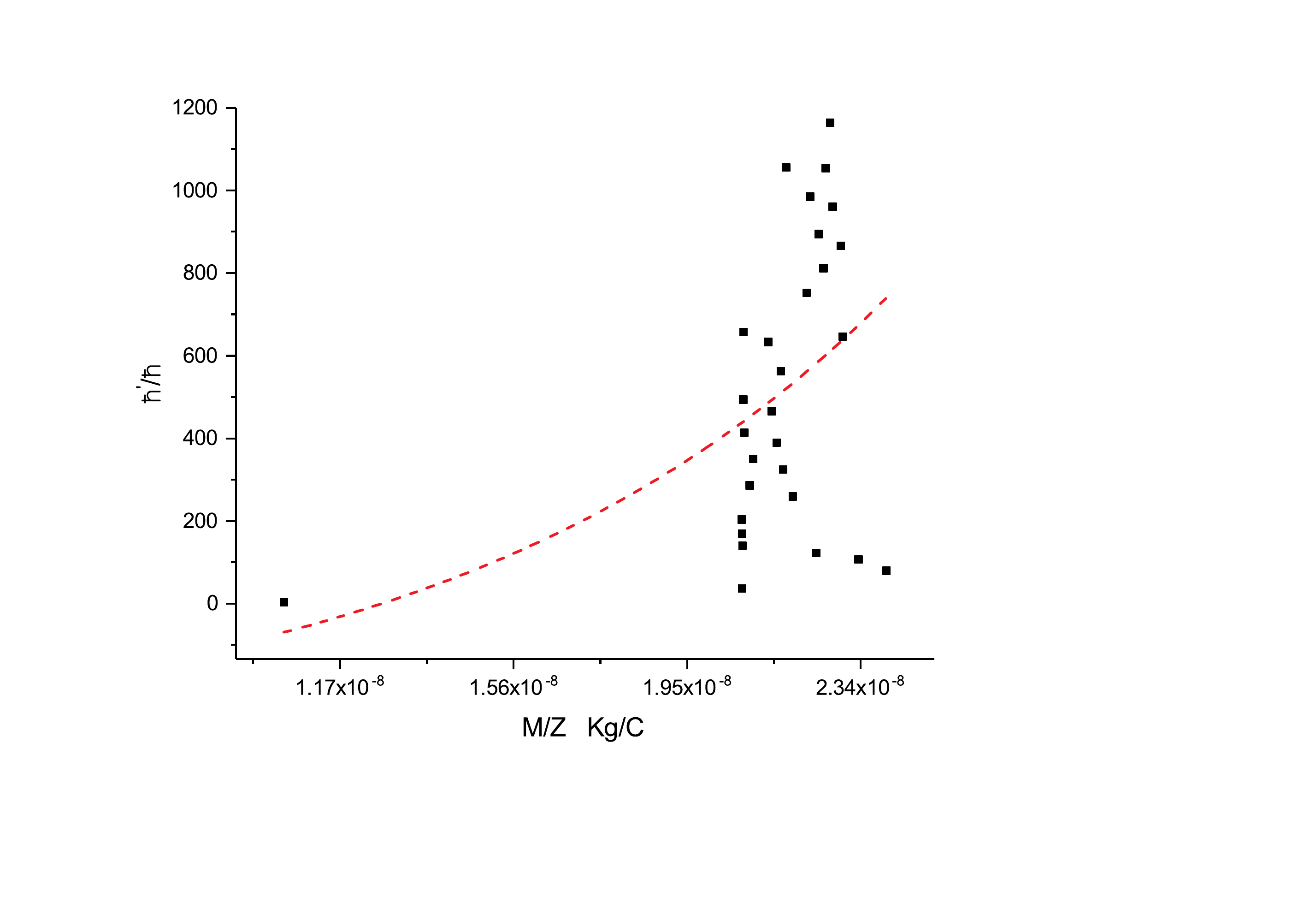}
\vspace{1pt}
\caption{ $\hbar^{\prime}/\hbar$ ratio versus $M/Z$ where $M$ is atomic mass and Z is atomic charge.The black points represent the calculated values of $\hbar^{\prime}/\hbar$ ratio, and the red dashed curve represents the fitting curve $\hbar^{\prime}/\hbar=0.03936~~\exp\left( \left(M/Z+ 7.878\times10^{-8}\right)/1.00425\times10^{-8}\right)- 353.486$, with the Coefficient of Determination $R^{2}= 0.15731$.}
\label{varyhhh1}
\end{figure}

\begin{figure}[htp]
\centering
\includegraphics[scale=0.4]{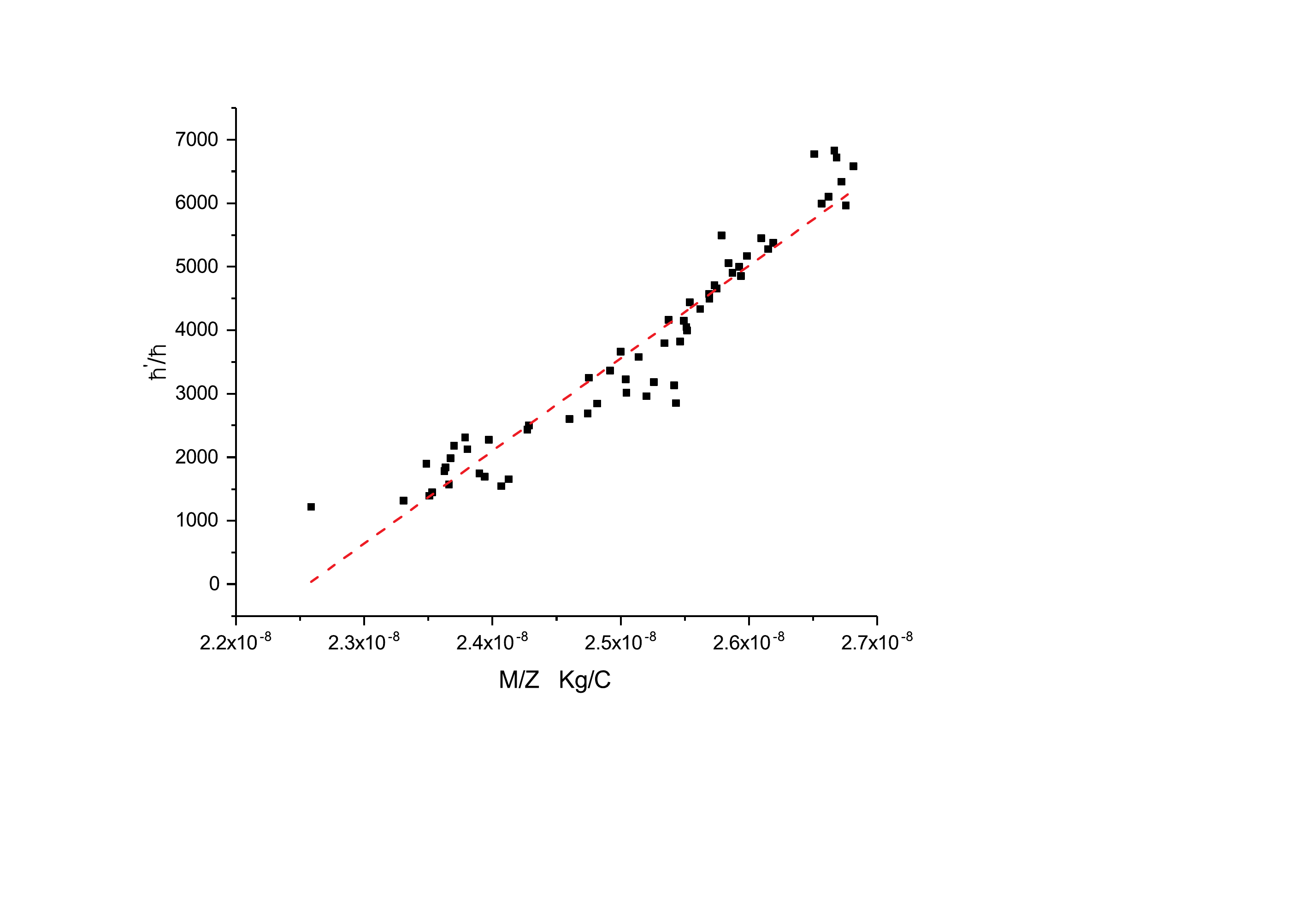}
\caption{$\hbar^{\prime}/\hbar$ ratio versus $M/Z$.The black points represent the calculated values of $\hbar^{\prime}/\hbar$ ratio, and the red dashed line represents the fitting curve $\hbar^{'}/\hbar= 1.45862\times10^{12} \dfrac{M}{Z}- 32908.49273$, with the Coefficient of Determination $R^{2}= 0.91925$.}
\label{varyhhh2}
\end{figure}

\begin{figure}[htp]
\centering
\includegraphics[scale=0.4]{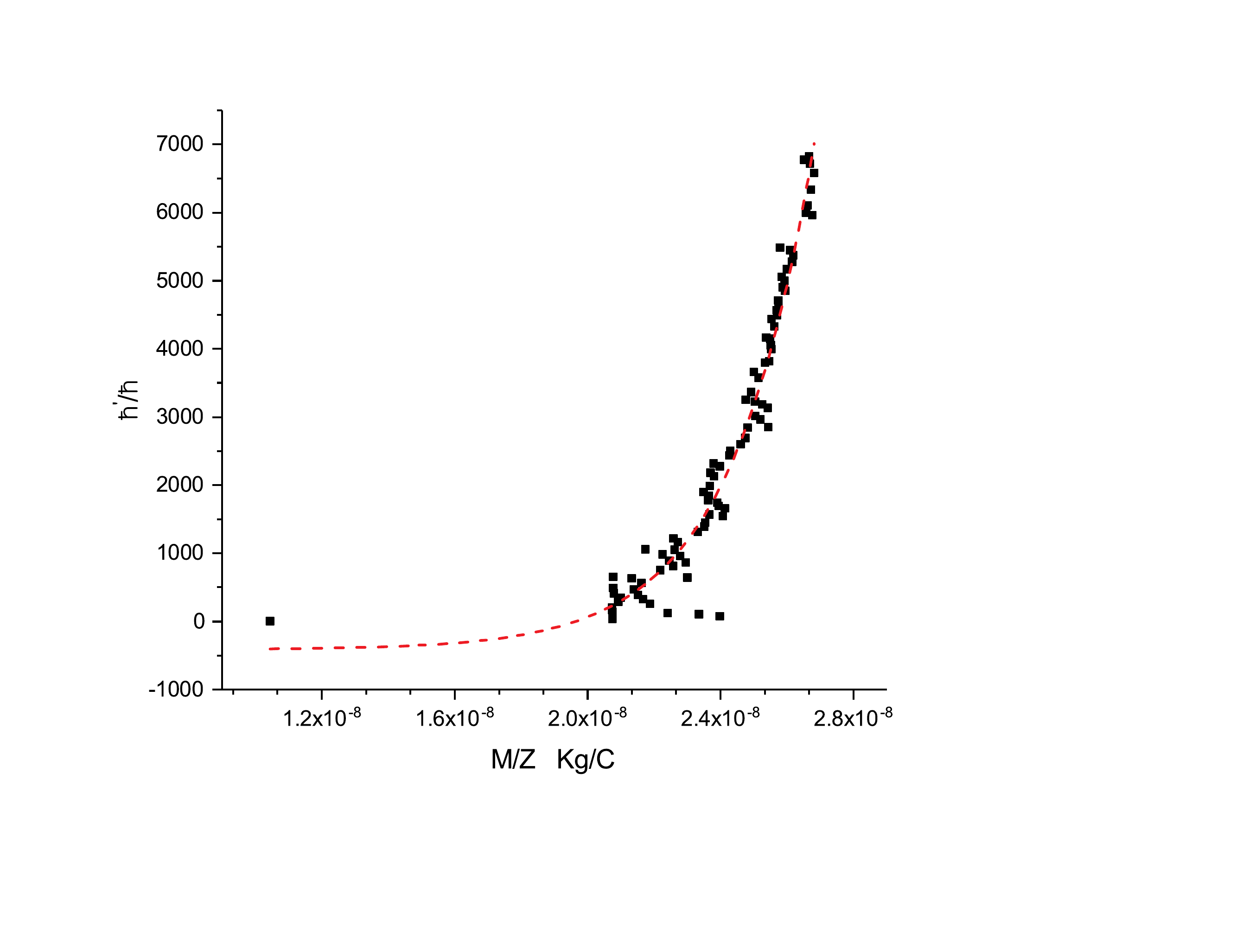}
\caption{ $\hbar^{\prime}/\hbar$ ratio versus  $M/Z$.The black points represent the calculated values of $\hbar^{\prime}/\hbar$ ratio, and the red dashed curve represents the fitting curve $\hbar^{\prime}/\hbar= 0.152 \exp \left(\frac{M/Z}{2.48346\times10^{-9}} \right)- 412.27738$, with the Coefficient of Determination $R^{2}= 0.95718$.}
\label{varyhhh3}
\end{figure}

We thus extend our argument to the nuclei of elements in the periodic table and their corresponding radii \cite{ANGELI201369}. All values are shown in data table in the appendix.  The relation between $\hbar^{\prime}/\hbar$ versus $M/Z$ is plotted in Figure~\ref{varyhhh1} for $Z<30$ and in Figure~\ref{varyhhh2} for $Z \ge 30$. In the former Figure, it is apparent the ratio $\hbar^{\prime}/h$ is fluctuating for small values of atomic mass $M$, while in the latter it starts to become linear for large values. The combined plots are collectively represented in Figure~\ref{varyhhh3} that shows approximately exponential behaviour for varying $\hbar$.

\begin{figure}[htp] 
\centering
\includegraphics[scale=0.4]{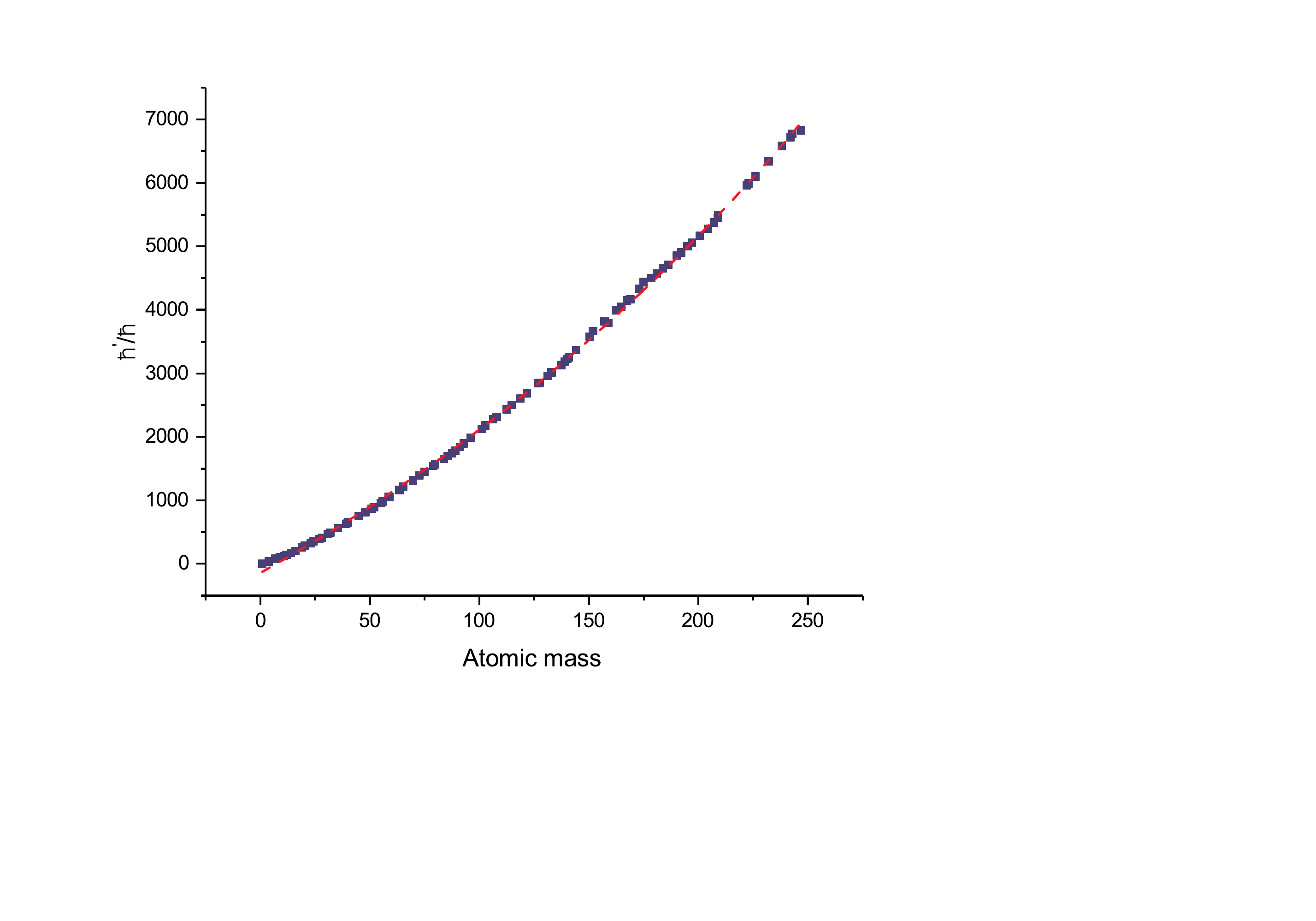}
\caption{ $\hbar^{\prime}/\hbar$ ratio versus the atomic mass $M$ of periodic table elements.The blue points represent the calculated values of $\hbar^{\prime}/\hbar$ ratio, and the red dashed curve represents the fitting curve $\hbar^{'}/\hbar= 6381.36102 \exp \left(M/330.51601\right)- 6522.91287$, with the Coefficient of Determination $R^{2}= 0.99949$.}
\label{varyhhh4}
\end{figure}

The fitting equation describing the relation shown in Figure~\ref{varyhhh1} is given by:
\begin{equation}\label{fitlin1}
   \hbar^{\prime}/\hbar= 0.03936 \exp \left(\frac{M/Z+ 7.878\times10^{-8}}{1\times10^{-8}} \right)- 353.48,
\end{equation}
while those for Figure~\ref{varyhhh2} and \ref{varyhhh3} are, respectively:
\begin{eqnarray}
\label{fitexp2}
\hbar^{\prime}/\hbar&=& 1.45\times10^{12} M/Z- 32908.49,\\ 
\label{fitlin3}
\hbar^{\prime}/\hbar&=& 0.15~\exp \left(\frac{M/Z}{2.48\times10^{-9}} \right)- 412.27,
\end{eqnarray}
The ratio between the effective Planck constant to the standard Planck's constant $\hbar^{\prime}/\hbar$ as a function of the atomic mass $M$ of elements is shown in Figure~\ref{varyhhh4} and it is described by the following equation:
\begin{equation}
    \hbar^{'}/\hbar= 6381.36 \exp \left(M/330.52\right)- 6522.91,
\label{fitexp4}
\end{equation}
This would tell us how the uncertainty principle loses its meaning by increasing the bound in the inequality of Equation~(\ref{varyh}). If one assumes $h^\prime$ eventually becomes infinite, this would imply that variance of momentum ($\Delta p$) or variance of position ($x$) would also be bounded by infinity.

Alternatively, one can postulate that a GUP-based variation of $\hbar$ may be connected with Dirac's Large Number Hypothesis, which itself implies a varying gravitational constant $G^\prime$ \cite{Dirac:1978xh,Dirac:1975vq}:
\begin{equation}
G ^{\prime} \propto 1/t \label{Dirac}~,
\end{equation}
which is supported by several astrophysical studies \cite{Barrow:1996xn,Barrow:2005hw,Christodoulou:2018xxw}. Conversely, we found in our analysis that the effective $\hbar^\prime$ is varying and increasing with increasing the complexity of matter, or simply with increasing atomic mass.  If complexity increases over the evolution of time \cite{Susskind:2018pmk}, it is logical to conclude that varying $\hbar^\prime$ is increasing with time evolution. This is shown in Figure~\ref{varyhhh3} and  Figure~\ref{varyhhh4}. In general $\hbar^\prime$ can be represented by a function of time as following:
\begin{eqnarray}
\hbar^{\prime} \propto f(t), \label{ourhypothesis}
\end{eqnarray}
Determining the exact  function of proportionality between $\hbar$ and $t$ is left for future investigation. We propose that it can be determined based determining the creation time of chemical elements based on Big bang nucleosynthesis  \cite{PhysRev.73.803,cyburt2016big}. By comparing Equation~(\ref{Dirac}) with Equation~ (\ref{ourhypothesis}), we find the following fundamental relation between the effective $\hbar^\prime$ and effective $G^\prime$ as follows,
\begin{eqnarray}
\hbar^{\prime} G^{\prime} &\propto& g(t),\label{QG} \nonumber\\ \hbar^{\prime}~G^{\prime}&=&\hbar~G~ g(t), \label{varyingtime}
\end{eqnarray}
where $g(t)=f(t)/t$, which we dub the quantum gravity complementary relation. It is obvious that when $g(t)\rightarrow1, \hbar^{\prime}~G^{\prime} \rightarrow \hbar~G~$.
Similar relation between $\hbar^{\prime}$ and $G^{\prime}$ was found in the timeless state of gravity  where it was found it is related to fifth power of distance measured from the gravitational source \cite{Ali:2021ela}. It is worth mentioning that a similar function to $g(t)$  was suggested in \cite{Basilakos:2010vs} as $f_{\pm}$ to introduce variation of Planck quantities due to GUP. 

One may additionally wonder how $\hbar^{\prime} G^{\prime}$ is defined in terms of geometry. To answer this question, we appeal to the Bekenstein-Hawking entropy area law \cite{Bekenstein:1973ur,Hawking:1975vcx} and note this contains the product $\hbar G$, {\it i.e.}

\begin{eqnarray}
S_{\text{BH}}= \frac{c^3 A}{4 G \hbar} \label{BH}
\end{eqnarray}
Recently, it was suggested that this relation should be modified by a Von Neumann entropy term in order to preserve the second law of thermodynamics, as well as to preserve information inside and outside the horizon \cite{Penington:2019npb,Maldacena:2020ady}, {\it i.e.}

\begin{eqnarray}
\label{SBH}
S_{\text{BH}}= \frac{ c^3 A_{\text{H}}}{4 G \hbar}+ S_{\text{matter}},
\end{eqnarray}
where Eq. (\ref{SBH}) can be rearranged  as follows:
\begin{eqnarray}
\hbar~G~\left(1-\frac{S_{\text{matter}}}{S_{\text{BH}}}\right)= \hbar^\prime G^\prime = \frac{ c^3 A_{\text{H}}}{4 S_{\text{BH}}}, \label{SBH1}
\end{eqnarray}
When comparing Eq. (\ref{varyingtime}) with Eq. (\ref{SBH1}), we obtain an expression for the function $g(t)$ that implies a time-dependent variation of both $\hbar^\prime$ and $G^\prime$:
\begin{eqnarray}
g(t)= 1-\frac{S_{\text{matter}}}{S_{\text{BH}}}, \label{equal}
\end{eqnarray}
For a pure state case in which $S_{\text{matter}}=0$ one finds $g(t)=1$, implying fixed values for $\hbar$ and $G$ according to Eq.~(\ref{varyingtime}). For the case of mixed states, {\it i.e} $S_{\text{matter}}\neq 0$, the expression implies a time dependence of $g(t)$. This may establish a correlation between mixed states in quantum mechanics and time as an emergent concept. In other words, time emerges from mixing states. Conversely, it may also be the case that a mixing of states is a result of the arrow of time.

Moreover, the fundamental function $g(t)$ may be a manifestation of the holographic principle between matter and information \cite{tHooft:1993dmi,Susskind:1994vu}. We further suggest that $g(t)$ explains why the  Von Neumann entropy correction is needed \cite{Penington:2019npb,Maldacena:2020ady} using our new perspective on varying $\hbar$ and $G$. Using Eq. (\ref{varyingtime}), we simply can write the effective variation of the Planck constant and gravitational constant  as follows:

\begin{eqnarray}
\hbar^{\prime} G^{\prime}=\hbar G \left(1-\frac{S_{\text{matter}}}{S_{\text{BH}}}\right), \label{solution}
\end{eqnarray}
\\
This forms a crucial bridge between the quantum and gravitational worlds, and discloses several interpretations. First, a varying $\hbar$ is found to complement Dirac's Large Number Hypothesis and varying $G$ \cite{Dirac:1978xh,Dirac:1975vq}. In addition, Equation (\ref{solution})  supports the thermodynamic explanation of gravity and matter \cite{Jacobson:1995ab,Verlinde:2010hp,Wald:1993nt,Verlinde:2016toy}. Implications  on classical and quantum information theory can be also investigated \cite{Witten:2018zva}.

In conclusion, we offer an outlook on a variety of additional novel implications of a varying $\hbar^\prime$. First, modifications of gravity should follow from Eq.~(\ref{solution}), from which a varying $G$. can be inferred \cite{Moffat:2007nj}. Furthermore, Eq.~(\ref{solution}) can be interpreted to suggest that the Von Neumann entropy term is itself the source of modified gravity.

A varying $\hbar^\prime$ could also replace the renormalization functions in quantum field theories, due to the fact that GUP or minimal length scenarios imply a natural cutoff scale that yields finite measurable values \cite{Chang:2001bm,Ali:2011ap,reuter2006minimal}. Since this cutoff ({\it i.e.} the minimal length) is included in the definition of $\hbar^\prime$, it may also resolve the renormalization problem in approaches to quantum gravity \cite{tHooft:2016fzb}. Recently, it was shown that entanglement entropy is varying in scattering processes \cite{Seki:2014cgq,Kharzeev:2017qzs}, which supports our proposal on the correlation between minimal length of scattering that is included in $\hbar^{\prime}$ and entropy. Lastly, it is expected that Equation~(\ref{solution}) could have implications in condensed matter systems where  the entropy-area law has wide applications \cite{Eisert:2008ur}.

As a final thought, it is worth mentioning that function $g(t)$ may be determined by comparing it with the element formation timeline in the context of big bang nucleosynthesis \cite{PhysRev.73.803,cyburt2016big,Luciano:2021vkl}. If we determine the form of $g(t)$ from the astrophysical observations and nucleosynthesis, we  expect that the  function $g(t)$ will  play a significant role in re-interpretation of renormalization functions in quantum field theories in terms of astrophysical data and black hole physics. We hope to report further investigations on these topics in the future.

\begin{acknowledgements}
Ahmed Farag Ali thanks Sahil Gupta for inspiring discussions around the connection between proton charge radius and Planck constant. J.M.~is a KITP Scholar at the Kavli Institute for Theoretical Physics. The KITP Scholars Program is supported in part by the National Science Foundation under Grant No.~NSF PHY-1748958.
\end{acknowledgements}

\bibliographystyle{utcaps}
\bibliography{ref21.bib}{}

\appendixpage
\renewcommand{\arraystretch}{1.4}
\begin{longtable}
[width= 9 cm]{ | l | l | l | l | l | l | }
\caption{Data Table} \label{tab:long} \\
\hline \scriptsize{Element} & \scriptsize{Z}& \scriptsize{Atomic weight}& $\frac{\text{Mass}}{10^{26}}$\scriptsize{Kg} & $\text{Charge radius}$~\;fm & $\dfrac{\hbar^{'}}{\hbar}$ \\

\hline \hline
 \hline H &1 &1.00797 & 0.16737 &0.8783 & 4.18 \\
 \hline He &2 &4.00260 & 0.6646 & 1.9661 &37.10\\
 \hline Li &3& 6.94100 &1.1526& 2.4440 & 80.10\\
 \hline Be &4 &9.01218 & 1.4965 & 2.5190 & 107.00\\
 \hline B&5 &10.81000 & 1.7950 &2.4060& 123.00\\
 \hline C &6 & 12.01100 & 1.9944 & 2.4702 & 140.00\\
 \hline N &7 & 14.00670 & 2.3258 & 2.5582 & 169.00\\
 \hline O &8 & 15.99940 & 2.6567 &2.6991 & 204.00\\
 \hline F &9 & 18.99840 & 3.1547 & 2.8976 & 260.00\\
 \hline Ne &10 & 20.17900 & 3.3507 & 3.0055 & 286.00\\
 \hline Na &11 & 22.98977 & 3.8175 & 2.9936 & 325.00\\
 \hline Mg &12 & 24.30500 & 4.0359 & 3.0570 & 351.00\\
 \hline Al &13 & 26.98154 & 4.4803 & 3.0610 & 390.00\\
 \hline Si & 14 & 28.08550 & 4.6636 & 3.1224 & 414.00\\
 \hline P &15 & 30.97376 & 5.1432 & 3.1889 & 466.00\\
 \hline S & 16 & 32.0600 & 5.3236 & 3.2611 & 494.00\\
 \hline Cl &17 & 35.45300 & 5.8869 &3.3654 & 563.00\\
 \hline Ar&18&39.94800&6.6334&3.4274& 646.00\\
 \hline K&19& 39.09830&6.49227 & 3.4349& 634.00\\
 \hline Ca & 20 & 40.0800 & 6.6553& 3.4776 & 658.00\\
 \hline Sc&21& 44.95590&7.4649 &3.5459 & 752.00\\
 \hline Ti &22& 47.90000&7.9538 & 3.5921 & 812.00\\
 \hline V &23& 50.94150 & 8.4589 & 3.6002 & 866.00\\
 \hline Cr&24& 51.99600 & 8.6339 & 3.6452 & 895.00\\
 \hline Mn&25 & 54.93800&9.1225&3.7057 & 961.00\\
 \hline Fe&26& 55.84700 & 9.2734& 3.7377 & 985.00\\
 \hline Co &27 & 58.93320 & 9.7859 &3.7875&1050.00\\
 \hline Ni &28& 58.70000 & 9.7471 & 3.8118 & 1060.00\\
 \hline Cu &\;29 & 63.54600 &10.5518 & 3.8823 & 1160.00\\
 \hline Zn &30& 65.38000&10.8563 & 3.9491 & 1220.00\\
 \hline Ga &31& 69.72000 & 11.5770 & 3.9998 &1320.00\\
 \hline Ge &32 & 72.59000 & 12.0536 & 4.0632 &1390.00\\
 \hline As &33& 74.92160 &12.4407 & 4.0968 & 1450.00\\
 \hline Se &34 & 78.96000& 13.1110& 4.1400 & 1540.00\\
 \hline Br &35& 79.90400 & 13.2681& 4.1629& 1570.00\\
 \hline Kr&36&83.80000&13.9150&4.1884& 1660.00\\
 \hline Rb&37&85.46780&14.1919&4.2036&1700.00\\
 \hline Sr&38&87.62000&14.5493& 4.2240 &1750.00\\
 \hline Y&39& 88.90590&14.7628 &4.2430&1780.00\\
 \hline Zr&40& 91.22000&15.1471&4.2845&1840.00\\
 \hline Nb&41 & 92.90640&15.4271&4.3240 &1900.00\\
 \hline Mo&42& 95.94000 &15.9308&4.3847 &1990.00\\
 \hline Ru &44 & 101.07000 &16.7827 &4.4606 & 2130.00\\
 \hline Rh&45 &102.90550 &17.0875&4.4945&2180.00\\
 \hline Pd &46& 106.40000&17.6677 & 4.5318 & 2280.00\\
 \hline Ag&47&107.86800 &17.9115&4.5454 &2310.00\\
 \hline Cd &48 & 112.41000 & 18.6657 &4.5944 &2440.00\\
 \hline In &49 &114.82000& 19.0659&4.6156 & 2500.00\\
 \hline Sn&50 &118.69000&19.7085&4.6438&2600.00\\
 \hline Sb &51 & 121.75000 &20.2166 &4.6802& 2690.00\\
 \hline Te &52 & 127.60000 & 21.1880 & 4.7346 &2850.00\\
 \hline I &53 &126.90450 &21.0725 & 4.7500 & 2850.00\\
 \hline Xe &54 & 131.30000 &21.8024 &4.7808 & 2960.00\\
 \hline Cs &55& 132.90540 &22.0689& 4.8041 & 3010.00\\
 \hline Ba &56& 137.33000 &22.8036 &4.8314 & 3130.00\\
 \hline La &57&138.90550& 23.0653 & 4.855&3180.00\\
 \hline Ce &58&140.12000 &23.2669 &4.8771 & 3230.00\\
 \hline Pr &59&140.90770& 23.3977&4.8919& 3250.00\\
 \hline Nd &60 &144.24000&23.9511 & 4.9421&3360.00\\
 \hline Sm &62&150.40000 & 24.9739 & 5.0387 & 3580.00\\
 \hline Eu &63& 151.96000 &25.2330 &5.1064 &3660.00\\
 \hline Gd &64 &157.25000 &26.1114 &5.1449 & 3820.00\\
 \hline Tb &65 & 158.92540 & 26.3896 & 5.0600 & 3800.00\\
 \hline Dy &66 & 162.50000 &26.9831 &5.2099 & 4000.00\\
 \hline Ho &67 & 164.93040 & 27.3867 &5.2022 & 4050.00\\
 \hline Er &68 &167.26000 &27.7735 &5.2560 & 4150.00\\
 \hline Tm &69 & 168.93420 & 28.0515& 5.2256 & 4170.00\\
 \hline Yb &70&173.04000 &28.7333 &5.3046 & 4330.00\\
 \hline Lu &71 &174.96700&29.0533 & 5.3700 &4440.00\\
 \hline Hf &72 &178.49000 & 29.6383 &5.3371 &4500.00\\
 \hline Ta &73 &180.94790 &30.0464 &5.3507 & 4570.00\\
 \hline W &74 &183.85000 & 30.5283 &5.3658 &4660.00\\
 \hline Re &75 &186.20700 &30.9197 &5.3596 & 4710.00\\
 \hline Os &76 &190.20000 & 31.5827 & 5.4062 & 4850.00\\
 \hline Ir &77 & 192.22000 &31.9180 &5.4032 & 4900.00\\
 \hline Pt &78 & 195.09000 & 32.3947 & 5.4270 & 5000.00\\ 
 \hline Au &79 &196.96650 &32.7063 & 5.4371 & 5060.00\\
 \hline Hg &80 &200.59000 &33.3080 & 5.4581 & 5170.00\\
 \hline TI &81 & 204.37000 & 3.9357 &5.4704 & 5280.00\\
 \hline Pb &82 & 207.20000 & 34.4056 &5.4943 &5370.00\\
 \hline Bi &83 & 208.98040  &34.7012 & 5.5211 & 5450.00\\
 \hline Po &84 &209.00000& 34.7045 & 5.5628&5490.00\\ \hline  Rn &86 & 222.00000 & 36.8631 & 5.6915 & 5960.00\\
 \hline Fr &87 & 223.00000 & 37.0292 & 5.6951 & 6000.00\\
 \hline Ra &88 & 226.02540 & 37.5315 & 5.7211& 6100.00\\
 \hline Th &90 & 232.03810 & 38.5299 & 5.7848 & 6340.00\\
 \hline U &92 & 238.02900 &39.5247 & 5.8571 & 6580.00\\
 \hline Pu &94 & 242.00000 & 40.1841 & 5.8823 & 6720.00\\
 \hline Am &95 & 243.00000  & 40.3502 & 5.9048 & 6770.00\\
 \hline Cm &96 & 247.00000& 41.0144& 5.8562& 6830.00\\
 \hline
\end{longtable}

\end{document}